\documentclass[%
twocolumn]{revtex4-2}

\usepackage[pdftex]{graphicx}
\usepackage{dcolumn}
\usepackage{bm}
\usepackage{xcolor}
\usepackage{amsmath}
\usepackage{braket}
\usepackage{appendix}
\usepackage{hyperref}

\begin{document}

\title{Magnetic octupole tensor decomposition and second-order magnetoelectric effect}

\author{Andrea Urru}
\affiliation{Department of Materials, ETH Zurich, CH-8093 Zürich, Switzerland} 
\author{Nicola A. Spaldin}
\affiliation{Department of Materials, ETH Zurich, CH-8093 Zürich, Switzerland}


\begin{abstract}

We discuss the second-order magnetoelectric effect, in which a quadratic or bilinear electric field induces a linear magnetization, in terms of the ferroic ordering of magnetic octupoles. We present the decomposition of a general rank-3 tensor into its irreducible spherical tensors, then reduce the decomposition to the specific case of the magnetic octupole tensor, $\mathcal{M}_{ijk} = \int \mu_i (\mathbf{r}) r_j r_k d^3 \mathbf{r}$. We use first-principles density functional theory to compute the size of the local magnetic multipoles on the chromium ions in the prototypical magnetoelectric Cr$_2$O$_3$, and show that, in addition to the well established local magnetic dipoles and magnetoelectric multipoles, the magnetic octupoles are non-zero. The magnetic octupoles in Cr$_2$O$_3$ have an anti-ferroic arrangement, so their {\it net} second-order magnetoelectric response is zero. Therefore they form a kind of hidden order, which could be revealed as a linear magnetic (antiferromagnetic) response to a non-zone-center (uniform) quadratic electric field.
\end{abstract}

\maketitle

\section{Introduction} 
\label{intro}

The linear magnetoelectric effect, in which an applied electric (magnetic) field induces a linear magnetization (electric polarization), was first observed in Cr$_2$O$_3$ in 1960 \cite{Astrov:1960}, following its prediction in the same year by Igor Dzyaloshinksii \cite{Dzyaloshinskii:1960}.  
Its existence requires terms of the form $\mathbf{H} \otimes \mathbf{E}$ (or $\alpha_{ij} H_i E_j$) in the free energy, where $\mathbf{H}$ and $\mathbf{E}$ are the magnetic and electric fields respectively, and so it can be non-zero only in materials in which both time-reversal and space-inversion symmetries are broken. It therefore occurs in the corresponding sub-set of antiferromagnets, which will be the focus of our discussion here, as well as in multiferroic materials that have spontaneous ferromagnetism and ferroelectricity.

The linear magnetoelectric effect has an obvious fundamental appeal in enabling modification of a material property using a handle other than its conjugate field. In addition, it is of potential technological interest in spintronic devices, since it allows control of magnetic properties using energy-efficient electric fields \cite{Binek/Doudin:2005}. It has been shown that magnetoelectric annealing in combined magnetic and electric fields can be used to select a specific antiferromagnetic domain \cite{Borisov_et_al:2005}, and that the surface of a linear magnetoelectric material always carries a net magnetic dipole moment \cite{Andreev:1996,Belashchenko:2010,Spaldin:2021}, which might in turn influence the exchange bias properties. 

A convenient formalism for describing linear magnetoelectric materials is provided by their magnetoelectric multipole tensor, ${\cal M}_{ij} = \int r_{i} {\mu_j(\mathbf{r})}d^3 \mathbf{r}$, which is the second-order coefficient in the multipole expansion of the energy, $E$, of a non-uniform magnetization, $\boldsymbol{\mu}(\mathbf{r})$, in a spatially varying magnetic field, $\mathbf{H}(\mathbf{r})$ \cite{Ederer/Spaldin:2007,Spaldin/Fiebig/Mostovoy:2008,Spaldin_et_al:2013}: 
\begin{eqnarray}
E & = & -  \int \boldsymbol{\mu}(\mathbf{r}) \cdot
\mathbf{H}(\mathbf{r}) \, d^3 \mathbf{r} \nonumber \\
 & = & - \int \boldsymbol{\mu}(\mathbf{r}) \cdot \mathbf{H}(0) \, d^3 \mathbf{r} 
-  \int \mu_{i} (\mathbf{r}) r_{j} \partial_{j} H_{i}(0) \, d^3 \mathbf{r} \quad  - \ldots \nonumber \\
\label{Eqn:dipole}
\end{eqnarray}
where the expansion in powers of the field gradients is calculated at some arbitrary reference point $\mathbf{r}
= 0$, and $i,j$ are Cartesian directions with summation over repeated indices implied. 

Of course, the expansion of Eq. \eqref{Eqn:dipole} can be continued to higher orders, with the coefficient in the next, third-order term, 
\begin{equation}
\label{Eqn:octupole}
- \int \mu_i (\mathbf{r}) r_{j} r_{k} \partial_{j} \partial_{k} H_{i}(0) \, d^3 \mathbf{r},
\end{equation}
containing a rank-three tensor of the form $ {\cal M}_{ijk} = \int  \mu_i (\mathbf{r}) r_j r_k d^3 \mathbf{r} $. The components of this tensor, conventionally referred to as magnetic octupoles, describe the second-order magnetoelectric effect in which a quadratic (or bilinear) electric field induces a linear magnetization, appearing as an $\mathbf{H} \otimes \mathbf{E} \otimes \mathbf{E}$ (or $\beta_{ijk} H_i E_j E_k$) term in the free energy. In contrast to the linear magnetoelectric multipole tensor, this tensor has non-zero components in materials that break time-reversal symmetry and are space-inversion symmetric. 

(Note that there is also a term of the form $\mathbf{H} \otimes \mathbf{H} \otimes \mathbf{E} $ at the same order in the free energy, corresponding to multipoles of the form $\mu_i \mu_j r_k$. These multipoles occur in materials that are time-reversal symmetric and break space-inversion symmetry, in which a quadratic or bilinear magnetic field induces a linear electric polarization. This term does not emerge from the multipole expansion of Eq. \eqref{Eqn:dipole} and we do not consider it here).

While the ${\cal M}_{ij}$ and ${\cal M}_{ijk}$ tensors contain all of the information about their respective multipoles, it is often useful to decompose them into their irreducible components. 

For the case of the magnetoelectric multipole tensor it is straightforward to show that these are the scalar magnetoelectric monopole \cite{Spaldin_et_al:2013}, 
\[a =  \frac{1}{3} {\cal M}_{ii} = \frac{1}{3} \int \boldsymbol{\mu}(\mathbf{r}) \! \cdot \mathbf{r} \, d^3 \mathbf{r} \quad , \] 
the toroidal moment vector \cite{Ederer/Spaldin:2007}, 
\[\mathbf{t} = - \frac{1}{2}  \int \boldsymbol{\mu}(\mathbf{r}) \! \times \mathbf{r} \, d^3 \mathbf{r}\] 
with components 
\[t_i = \frac{1}{2} \varepsilon_{ijk} {\cal M}_{jk} \quad , \] 
and the traceless quadrupole tensor, 
\begin{eqnarray}
q_{ij} &=& \frac{1}{2}\left({\cal M}_{ij} + {\cal M}_{ji} - \frac{2}{3} \delta_{ij} {\cal
M}_{kk}\right)\nonumber\\
&= &\frac{1}{2} \int \left[r_i \mu_j +
r_j \mu_i - \frac{2}{3} \delta_{ij} \mathbf{r}\! \cdot \boldsymbol{\mu}(\mathbf{r})
\right] d^3\mathbf{r} \quad ,
\end{eqnarray} 
with
${\cal M} = $
\begin{equation}
\label{eq_me_tensor}
\begin{pmatrix}
a + \frac{1}{2}q_{x^2-y^2} - \frac{1}{2} q_{z^2} 
& t_z + q_{xy} & t_y + q_{xz}\\
-t_z + q_{xy} & a - \frac{1}{2}q_{x^2-y^2} - \frac{1}{2} q_{z^2}  & -t_x + q_{yz} \\
-t_y + q_{xz} & t_x + q_{yz} &  a + q_{z^2}
\end{pmatrix}
.
\end{equation}

A linear magnetoelectric antiferromagnet can then be understood in terms of the ferroic ordering of the appropriate constituent magnetoelectric multipoles, which is more aesthetically appealing than considering the {\it anti-}ferroic ordering of its magnetic dipoles, $\mathbf{m} = \int \boldsymbol{\mu}(\mathbf{r}) d^3 \mathbf{r}$, which form the first term of the expansion of Eq. \eqref{Eqn:dipole}. 
The toroidal moment has proved to be particularly conceptually helfpul. Since it is a vector (compared with the scalar monopole and the quadrupole tensor), it provides a convenient time-reversal and space-inversion symmetry-breaking analogue to the magnetic and electric dipole moments, which break only time-reversal or space-inversion symmetries respectively. To our knowledge the toroidal moment was first introduced in 1958 in the context of parity violation by weak interactions by Zel’dovich, who pointed out its different interaction with a magnetic field from that of the magnetic dipole and quadrupole moments and named it an anapole moment \cite{Zeldovich:1958}. In condensed matter it has been proposed to underlie superdiamagnetic behavior \cite{Ginzburg_et_al:1984}, invoked to explain anomalies at magnetoelectric phase transitions \cite{Gorbatsevich/Kopaev/Tugushev:1983} and suggested as a possible ``hidden'' order parameter \cite{Chatel/Buin:2002}. Toroidal moments have been detected using resonant magnetoelectric X-ray scattering \cite{Arima_et_al:2005} and optical second harmonic generation has been used to image ferrotoroidal domains \cite{VanAken_et_al:2007}. 

Likewise, it is convenient to decompose the magnetic octupole tensor into its irreducible parts. In this case, these are a totally symmetric, traceless, tensor parametrized by the seven spherical components of the usual magnetic octupoles, and a residue non-symmetric tensor, whose entries in turn can be written in terms of two multipoles that we identify as the moment and the quadrupole of the toroidal moment density, defined as $\boldsymbol{\tau} (\mathbf{r}) = \mathbf{r} \times \boldsymbol{\mu} (\mathbf{r})$.
Ferroic ordering of these irreducible components can then lead to the second-order magnetoelectric effect mentioned above, as well as other second-order magnetic responses such as piezomagnetism and magnetostriction \cite{Patri_et_al:2019}. 
Magnetic octupoles are proposed as the {\it primary} order parameter for the phase transition in $5d^2$ double perovskites osmates, at which time-reversal symmetry breaking is indicated by muon spin resonance spectroscopy but neutron spectroscopy and diffraction find no magnetic dipoles \cite{Maharaj_et_al:2020}. And when geometrically frustrated by their arrangement in a crystal they can exhibit quantum liquid behavior analogous to the spin liquids proposed for magnetic dipoles \cite{Sibille_et_al:2020}. Moreover, they have been reported to be responsible for large magneto-optical Kerr rotation and for anomalous Hall effect in Mn$_3$Sn \cite{MOKE_octupoles, cluster_octupoles_Arita}. 

The remainder of this article is organized as follows. In the next section we provide a detailed derivation of the decomposition of the rank-3 magnetic octupole tensor into its irreducible parts, since, while this procedure is well established in the high-energy physics community, it is less familiar to many condensed-matter physicists. In Section \ref{sec3} we discuss the second-order magnetoelectric response in the prototype linear magnetoelectric Cr$_2$O$_3$. Finally, Section \ref{sec4} contains a summary of our findings.

\section{Magnetic Octupole decomposition}

\subsection{Decomposition of rank-3 tensors into irreducible spherical tensors}
\label{decomposition}

We consider a rank-3 tensor, written as the tensor product of three vectors, $\mathbf{a}$, $\mathbf{b}$, and $\mathbf{c}$, i.e. $M = \mathbf{a} \otimes \mathbf{b} \otimes \mathbf{c}$: the entries of $M$ are therefore $M_{ijk} = a_i b_j c_k$. It is common to decompose $M$ under the permutation group: 
\begin{equation}
\label{eq17}
    M_{ijk} = S_{ijk} + A_{ijk} + N_{ijk},
\end{equation}
where $S$ ($A$) is totally symmetric (totally anti-symmetric) with respect to exchange of any pair of indices, whereas $N$ is the \textit{residue tensor}, which in the general case is neither symmetric nor anti-symmetric on exchange of indices. Equivalently, $M$ can be written in terms of irreducible spherical tensors by decomposing it into irreducible representations (irreps) of SO(3). Since each of the vectors $\mathbf{a}$, $\mathbf{b}$, $\mathbf{c}$ transforms as a three-dimensional irrep of SO(3), the full tensor $M$ transforms as a $\mathbf{3} \otimes \mathbf{3} \otimes \mathbf{3}$ reducible representation, which decomposes into irreps in the following way:  
\begin{equation}
\label{eq12}
\begin{split}
    \mathbf{3} \otimes \mathbf{3} \otimes \mathbf{3} &= \mathbf{3} \otimes (\mathbf{5} \oplus \mathbf{3} \oplus \mathbf{1}) \\
    &= \underbrace{\mathbf{3} \otimes \mathbf{5}}_{=\mathbf{7} \oplus \mathbf{5} \oplus \mathbf{3}} \oplus \underbrace{\mathbf{3} \otimes \mathbf{3}}_{=\mathbf{5} \oplus \mathbf{3} \oplus \mathbf{1}} \oplus \underbrace{\mathbf{3} \otimes \mathbf{1}}_{=\mathbf{3}} \\
    &= \mathbf{7} \oplus \mathbf{5} \oplus \mathbf{5} \oplus \mathbf{3} \oplus \mathbf{3} \oplus \mathbf{3} \oplus \mathbf{1}.
\end{split}
\end{equation}

Below we review how the decompositions in Eqs. \eqref{eq17} and \eqref{eq12} are linked to each other, what each irreducible spherical tensor corresponds to and how they are built from the entries $M_{ijk}$. 

\subsubsection{Totally symmetric tensor}
The totally symmetric tensor $S$, defined as: 
\begin{equation}
\label{eq13}
S_{ijk} = \frac{1}{3!} \left[ M_{ijk} + M_{jki} + M_{kij} + M_{jik} + M_{kji} + M_{ikj} \right]
\end{equation}
has 10 independent components, and it can be further decomposed by removing the trace tensor. The trace of $S$: 
\begin{equation}
\label{eq18}
    \chi_i = S_{i j' k'} \delta_{j' k'},
\end{equation}
defined as the contraction of $S$ with a Kr\"onecker $\delta$, has three independent components ($i=1,2,3$), hence it transforms as a 3-dimensional irrep, i.e. as a vector. $\chi_i$ can be lifted to the rank-3 form as: 
\begin{equation}
\label{eq19}
    R_{ijk} = \chi_i \delta_{jk} + \chi_j \delta_{ik} + \chi_k \delta_{ij}, 
\end{equation}
where the three terms ensure that $R$ is totally symmetric. 

As a result, the remaining totally symmetric, traceless, tensor $T$, obtained as: 
\begin{equation}
T_{ijk} = S_{ijk} - \frac{1}{5} R_{ijk}, 
\end{equation}
where the factor $1/5$ enforces the trace of $T$ to be zero, transforms as a 7-dimensional irrep. Therefore, $S$ accounts overall for the $\mathbf{7} \oplus \mathbf{3}$ part of the decomposition reported in Eq. \eqref{eq12}. 

\subsubsection{Totally anti-symmetric tensor}
The totally anti-symmetric tensor $A$ has only one independent component, hence it transforms as a scalar and accounts for the irrep $\mathbf{1}$ in the decomposition provided in Eq. \eqref{eq12}. The scalar, identified as $\xi$ in the following, is obtained by contracting $M$ with the totally anti-symmetric Levi-Civita tensor as 
\begin{equation}
\label{eq15}
    \xi = \frac{1}{3!} M_{i'j'k'} \epsilon_{i'j'k'}, 
\end{equation}
and the tensor $A$, in turn, is written as: 
\begin{equation}
\label{eq38}
    A_{ijk} = \xi \epsilon_{ijk}.
\end{equation}

\subsubsection{Residue tensor}
Finally, the non-symmetric part $N$ accounts for the remaining 16 independent components. We show that it can be decomposed into the remaining irreps in Eq. \eqref{eq12}, i.e. $\mathbf{5} \oplus \mathbf{3} \oplus \mathbf{5} \oplus \mathbf{3}$.

The remaining invariant sub-tensors are obtained by contracting two indices of the parent tensor $M$ with the Levi-Civita tensor $\epsilon$. There are three possibilities for choosing the two indices to contract:
\begin{equation}
\begin{split}
    L_{ii'} &= M_{i j' k'} \epsilon_{j' k' i'}\\
    P_{ii'} &= M_{k' i j'} \epsilon_{j' k' i'} \\
    Q_{ii'} &= M_{j' k' i} \epsilon_{j' k' i'}.
\end{split}
\end{equation}
The resulting object has two free indices and transforms accordingly as a matrix, which is s a $\mathbf{5} \oplus \mathbf{3} \oplus \mathbf{1}$ reducible representation of SO(3). 

The matrices $L$, $P$, and $Q$ have the same trace: 
\begin{equation}
\label{eq34}
\begin{split}
     \chi_L &= L_{ii'} \delta_{ii'} = M_{i'j'k'} \epsilon_{i'j'k'} = 6 \xi, \\
     \chi_P &= P_{ii'} \delta_{ii'} = M_{k'i'j'} \epsilon_{i'j'k'} = 6 \xi, \\
     \chi_Q &= Q_{ii'} \delta_{ii'} = M_{j'k'i'} \epsilon_{j'k'i'} = 6 \xi, 
\end{split}
\end{equation}
which is proportional to the scalar $\xi$. Since $\xi$ is already accounted for in the totally anti-symmetric tensor transforming as the irrep $\mathbf{1}$, the trace of $L$, $P$, and $Q$ can be removed to redefine the traceless matrices $\widetilde{L}$, $\widetilde{P}$, and $\widetilde{Q}$: 
\begin{equation}
\begin{split}
\label{eq31}
\widetilde{L}_{ii'} &= L_{ii'} - 2 \xi \delta_{ii'}, \\
\widetilde{P}_{ii'} &= P_{ii'} - 2 \xi \delta_{ii'}, \\
\widetilde{Q}_{ii'} &= Q_{ii'} - 2 \xi \delta_{ii'}.
\end{split}
\end{equation}
The sum of these new matrices equals zero (see Appendix \ref{lin_dep} for more details): 
\begin{equation}
    \widetilde{L}_{ii'} + \widetilde{P}_{ii'} + \widetilde{Q}_{ii'} = 0,
\end{equation}
which means that $\widetilde{L}$, $\widetilde{P}$, and $\widetilde{Q}$ are linearly dependent. The choice of the two independent matrices is free and does not affect the results: in the following, we will take $\widetilde{L}$ and $\widetilde{P}$ as the independent ones. 

Each of the two independent matrices can be further decomposed into its irreducible components. Since both $\widetilde{L}$ and $\widetilde{P}$ are traceless, each of them transforms as a $\mathbf{5} \oplus \mathbf{3}$ reducible representation, and therefore they account for the remaining $\mathbf{5} \oplus \mathbf{3} \oplus \mathbf{5} \oplus \mathbf{3}$ part of the decomposition of $M$. 

The two matrices $\widetilde{L}$ and $\widetilde{P}$ accounting for the irreps $\mathbf{5} \oplus \mathbf{3} \oplus \mathbf{5} \oplus \mathbf{3}$, can be lifted to rank-3 tensors (see Appendix \ref{residue} for a more detailed derivation) in the following way: 
\begin{equation}
\label{eq14}
    \begin{split}
        N^1_{i j k} &= \frac{1}{3} \left( \widetilde{L}_{i i'} \epsilon_{i' j k} - \widetilde{L}_{k i'} \epsilon_{i' i j} \right), \\
        N^2_{i j k} &= - \frac{1}{3} \left( \widetilde{P}_{i i'} \epsilon_{i' j k} - \widetilde{P}_{j i'} \epsilon_{i' k i} \right),
    \end{split}
\end{equation}
and the full residue tensor is given by 
\begin{equation}
    N_{ijk} = N^1_{ijk} + N^2_{ijk}.
\end{equation}

\subsection{The specific case of the magnetic octupole tensor} 
\label{octupole_decomposition}

Next, we apply our derivation of Section \ref{decomposition} to the specific case of the rank-3 magnetic octupole tensor 
\begin{equation}
    \mathcal{M}_{ijk} = \int \mu_i (\mathbf{r}) r_j r_k d^3 \mathbf{r}, 
\end{equation}
which appears in the integral of Eq. \eqref{Eqn:octupole}. 

The magnetic octupole tensor is symmetric on exchange of $j$ and $k$, which reduces the number of independent parameters of $\mathcal{M}$ from the $27$ expected in the general case. $\mathcal{M}$ can be written as the tensor product of the vector with entries $\mu_i$, the magnetization density components, and the matrix with entries $r_j r_k$: the former has three linearly independent components, the latter has $6$ free parameters, being totally symmetric by exchange of $j$ and $k$. As a consequence, $\mathcal{M}$ has $3 \times 6 = 18$ independent parameters, i.e. it transforms as an $18$-dimensional reducible representation of SO(3), which we will show can be decomposed into irreps as: 
\begin{equation}
\label{eq27}
\begin{split}
    \mathbf{3} \otimes \left( \mathbf{3} \otimes \mathbf{3} \right)_{\text{symm}} &= \mathbf{3} \otimes (\mathbf{5} \oplus \mathbf{1}) \\
    &= \underbrace{\mathbf{3} \otimes \mathbf{5}}_{=\mathbf{7} \oplus \mathbf{5} \oplus \mathbf{3}} \oplus \underbrace{\mathbf{3} \otimes \mathbf{1}}_{=\mathbf{3}} \\
    &= \mathbf{7} \oplus \mathbf{5} \oplus \mathbf{3} \oplus \mathbf{3}.
\end{split}
\end{equation}
Compared to the general case, the scalar irrep $\mathbf{1}$, corresponding to the totally anti-symmetric tensor, is missing, hence $\xi = 0$ and Eq. \eqref{eq17} in this case reads: 
\begin{equation}
    \mathcal{M}_{ijk} = S_{ijk} + N_{ijk}.
\end{equation}

\subsubsection{Totally symmetric tensor, \texorpdfstring{$S_{ijk}$}{}} 
Since $\mathcal{M}_{ijk} = \mathcal{M}_{ikj}$, the totally symmetric part, $S_{ijk}$, defined in Eq. \eqref{eq13}, reads: 
\begin{equation}
\label{eq26}
    S_{ijk} = \frac{1}{3} \left( \mathcal{M}_{ijk} + \mathcal{M}_{jki} + \mathcal{M}_{kij} \right).
\end{equation}
$S$ transforms as $\mathbf{7} \oplus \mathbf{3}$; below, we further decompose it into the trace part ($\mathbf{3}$) and the totally symmetric, traceless part ($\mathbf{7}$).

\paragraph{Trace tensor} 
The traces of $S$ follow from Eq. \eqref{eq18} and, after some manipulation, read
\begin{align}
    \chi_1 &= \frac{1}{3} \int \left[ r^2 \mu_x (\mathbf{r}) + 2 (\boldsymbol{\mu} (\mathbf{r}) \cdot \mathbf{r}) x \right] d^3 \mathbf{r}, \\
    \chi_2 &= \frac{1}{3} \int \left[ r^2 \mu_y (\mathbf{r}) + 2 (\boldsymbol{\mu} (\mathbf{r}) \cdot \mathbf{r}) y \right] d^3 \mathbf{r}, \\ 
    \chi_3 &= \frac{1}{3} \int \left[ r^2 \mu_z (\mathbf{r}) + 2 (\boldsymbol{\mu} (\mathbf{r}) \cdot \mathbf{r}) z \right] d^3 \mathbf{r},
\end{align}
which can be cast into 
\begin{equation}
    \boldsymbol{\chi} = \frac{1}{3} \int \left[ r^2 \boldsymbol{\mu} (\mathbf{r}) + 2 (\boldsymbol{\mu} (\mathbf{r}) \cdot \mathbf{r}) \mathbf{r} \right] d^3 \mathbf{r}.
\end{equation}
The trace tensor $R$ introduced in Eq. \eqref{eq19} (and including already the factor $1/5$ in front of it), written in terms of the three independent parameters $\chi_1$, $\chi_2$, and $\chi_3$ reads 
\begin{align*}
    R_{1bc} &= \frac{1}{5} \begin{pmatrix} 3 \chi_1 & \chi_2 & \chi_3 \\ \chi_2 & \chi_1 & 0 \\ \chi_3 & 0 & \chi_1 \end{pmatrix}, \\
    R_{2bc} &= \frac{1}{5} \begin{pmatrix} \chi_2 & \chi_1 & 0 \\ \chi_1 & 3 \chi_2 & \chi_3 \\ 0 & \chi_3 & \chi_2 \end{pmatrix}, \\
    R_{3bc} &= \frac{1}{5} \begin{pmatrix} \chi_3 & 0 & \chi_1 \\ 0 & \chi_3 & \chi_2 \\ \chi_1 & \chi_2 & 3 \chi_3 \end{pmatrix}. 
\end{align*}

\paragraph{Totally symmetric, traceless tensor: magnetic octupoles}
\label{octupoles}
The entries of the totally symmetric, traceless, tensor $T$ can be computed from $S$ and the trace tensor $R$ as $T_{ijk} = S_{ijk} - R_{ijk}$. $T$ transforms as the irrep $\mathbf{7}$, hence its entries can be expressed in terms of $7$ independent components, the magnetic octupoles, which can be built from the spherical harmonics with $l = 3$ (see Appendix \ref{app_octupoles} for a more detailed discussion). These are 
\begin{widetext}
\begin{equation}
\label{eq25}
\begin{split}
    O_{-3} &= \frac{1}{3} \int \left[ \mu_y (\mathbf{r}) (3 x^2 - y^2) + 2 y (3 \mu_x (\mathbf{r}) x - \mu_y (\mathbf{r}) y) \right] d^3 \mathbf{r}, \\
    O_{-2} &= \frac{1}{3} \int \left[ \mu_x (\mathbf{r}) y z + \mu_y (\mathbf{r}) x z + \mu_z (\mathbf{r}) x y \right] d^3 \mathbf{r}, \\
    O_{-1} &= \frac{1}{3} \int \left[ \mu_y (\mathbf{r}) (4 z^2 - x^2 - y^2) + 2 y (4 z \mu_z (\mathbf{r}) - x \mu_x (\mathbf{r}) - y \mu_y (\mathbf{r})) \right] d^3 \mathbf{r}, \\
    O_{0} &=  \frac{1}{3} \int \left[ \mu_z (\mathbf{r}) (2 z^2 - 3 x^2 - 3 y^2) + 2 y (2 z \mu_z (\mathbf{r}) - 3 x \mu_x (\mathbf{r}) - 3 y \mu_y (\mathbf{r})) \right] d^3 \mathbf{r}, \\
    O_{1} &=  \frac{1}{3} \int \left[ \mu_x (\mathbf{r}) (4 z^2 - x^2 - y^2) + 2 x (4 z \mu_z (\mathbf{r}) - x \mu_x (\mathbf{r}) - y \mu_y (\mathbf{r})) \right] d^3 \mathbf{r}, \\
    O_{2} &=  \frac{1}{3} \int \left[ \mu_z (\mathbf{r}) (x^2 - y^2) + 2 z (x \mu_x (\mathbf{r}) - y \mu_y (\mathbf{r})) \right] d^3 \mathbf{r}, \quad \text{and}\\
    O_{3} &=  \frac{1}{3} \int \left[ \mu_x (\mathbf{r}) (x^2 - 3 y^2) + 2 x (x \mu_x (\mathbf{r}) - 3 y \mu_y (\mathbf{r})) \right] d^3 \mathbf{r}.
\end{split}
\end{equation}
\end{widetext}

The entries $T_{ijk}$ can then be written as linear combinations of the magnetic octupoles with appropriate coefficients, which can be determined by inspection. As an example, here we explicitly discuss $T_{(112)}$ (the round brackets identify all the possible permutations of the indices). After building $R_{(112)}$ and $S_{(112)}$ from Eqs. \eqref{eq19} and \eqref{eq26}, respectively, $T_{(112)}$  reads 
\begin{equation}
\label{eq28}
\begin{split}
    T_{(112)} = & \frac{1}{15} \int \left[ 4 \mu_y (\mathbf{r}) x^2 - 3 \mu_y (\mathbf{r}) y^2 - \mu_y (\mathbf{r}) z^2 \right. \\ & \left. + 8 \mu_x (\mathbf{r}) x y - 2 \mu_z (\mathbf{r}) y z \right] d^3 \mathbf{r},
\end{split}
\end{equation}
which can be written in terms of the octupoles $O_{-3}$ and $O_{-1}$ as
\begin{equation}
    T_{(112)} = \frac{1}{20} \left( 5 O_{-3} - O_{-1} \right). 
\end{equation}
By working out the same procedure for every entry, the full, totally symmetric, traceless tensor $T$ reads: 
\begin{widetext}
\begin{equation}
\label{eq35}
    \begin{split}
        T_{1jk} &= \frac{1}{20} \begin{pmatrix} 5 O_3 - 3 O_1 & 5 O_{-3} - O_{-1} & 2 \left( 5 O_2 - O_0 \right) \\ 
                                            5 O_{-3} - O_{-1} & - \left( 5 O_3 + O_1 \right) & 20 O_{-2} \\ 
                                            2 \left( 5 O_2 - O_0 \right) & 20 O_{-2} & 4 O_1 \end{pmatrix}, \\
        T_{2jk} &= \frac{1}{20} \begin{pmatrix} 5 O_{-3} - O_{-1} & -\left( 5 O_{3} + O_{1} \right) & 20 O_{-2} \\ 
                                            - \left( 5 O_{3} + O_{1} \right) & - \left( 5 O_{-3} + 3 O_{-1} \right) & - 2 \left( 5 O_2 + O_0 \right) \\ 
                                            20 O_{-2} & - 2 \left( 5 O_2 + O_0 \right) & 4 O_1 \end{pmatrix}, \\
        T_{3jk} &= \frac{1}{20} \begin{pmatrix} 2 \left( 5 O_2 - O_0 \right) & 20 O_{-2} & 4 O_1 \\ 20 O_{-2} & -2 \left( 5 O_2 + O_0 \right) & 4 O_{-1} \\ 4 O_1 & 4 O_{-1} & 4 O_0 \end{pmatrix}.
    \end{split}
\end{equation}
\end{widetext}

\subsubsection{Residue tensor, \texorpdfstring{$N_{ijk}$}{}}
In this case, the residue $N$ is symmetric on exchange of $j$ and $k$ because it must have the same symmetries as the parent tensor, $\mathcal{M}$. Of course, $N$ transforms as the remaining $\mathbf{5} \oplus \mathbf{3}$ reducible representation in Eq. \eqref{eq27}. This representation is smaller than the $\mathbf{5} \oplus \mathbf{3} \oplus \mathbf{5} \oplus \mathbf{3}$ we discussed for $N$ in the general case, due to the additional symmetry on exchange of $j$ and $k$. This is then responsible for the vanishing of one of the two tensors $N^1$ or $N^2$; by adopting the choice introduced in the previous section and discussed in Appendix \ref{residue}, $N^1$ must clearly vanish because it does not have the correct symmetry by permutation. This fact can be equivalently explained by considering the sub-tensors $\widetilde{L}$ and $\widetilde{P}$: their definition in Eq. \eqref{eq34} implies that $\widetilde{L} = 0$ since it is given by a contraction of a symmetric and an anti-symmetric tensor by permutation of $j$ and $k$. In turn, $N^1 = 0$ as it is defined in terms of $\widetilde{L}$ only. The only non-zero matrix is therefore $\widetilde{P}$, which we build entry by entry, following Eq. \eqref{eq34}. For instance, the entry $i=1$, $i'=1$ reads 
\begin{equation}
\begin{split}
    \widetilde{P}_{11} = \mathcal{M}_{312} - \mathcal{M}_{213} &= \int \left[ \mu_z (\mathbf{r}) x y - \mu_y (\mathbf{r}) x z \right] d^3 \mathbf{r}\\
    &= \int \left[ x (\mu_z (\mathbf{r}) y - \mu_y (\mathbf{r}) z) \right] d^3 \mathbf{r} \\
    &= \int x \tau_x (\mathbf{r}) d^3 \mathbf{r},
\end{split}
\end{equation}
where in the last line we used the definition of the toroidal moment density ($\boldsymbol{\tau} (\mathbf{r}) = \mathbf{r} \times \boldsymbol{\mu} (\mathbf{r})$). After computing all the remaining entries, $\widetilde{P}_{ij}$ reads 
\begin{equation}
    \widetilde{P}_{ij} = \int r_i \tau_j (\mathbf{r}) d^3 \mathbf{r},
\end{equation}
hence it has a form similar to the magnetoelectric multipole tensor discussed in Section \ref{intro}, the main difference being that the magnetization density $\boldsymbol{\mu} = (\mu_x, \mu_y, \mu_z)$ is replaced by the toroidal moment density $\boldsymbol{\tau} = (\tau_x, \tau_y, \tau_z)$. As a result, the decomposition of $\widetilde{P}$ into the irreps $\mathbf{5}$ and $\mathbf{3}$ is the same as for the magnetoelectric multipole tensor, with the substitution $(\mu_x, \mu_y, \mu_z) \rightarrow (\tau_x, \tau_y, \tau_z)$. Thus, the totally anti-symmetric part is 
\begin{equation}
    \widetilde{P}^{(A)} = \begin{pmatrix} 0 & t_z^{(\tau)} & -t_y^{(\tau)} \\ -t_z^{(\tau)} & 0 & t_x^{(\tau)} \\ t_y^{(\tau)} & -t_x^{(\tau)} & 0 \end{pmatrix},
\end{equation}
where $\mathbf{t}^{(\tau)} = \int \mathbf{r} \times \boldsymbol{\tau} (\mathbf{r}) d^3 (\mathbf{r})$ is the moment of the toroidal moment density and corresponds, in the spherical irreducible tensor notation $w^{k p r}$ (discussed in Ref. \cite{multipole_decomposition}), to $w^{2 1 1}$: 
\begin{align*}
    t_y^{(\tau)} &= \frac{i}{\sqrt{2}} \left( w^{211}_{-1} + w^{211}_{1} \right) = \int \left[ z \tau_x (\mathbf{r}) - x \tau_z (\mathbf{r}) \right] d^3 \mathbf{r}, \\
    t_z^{(\tau)} &= w^{211}_{0} = \int \left[ x \tau_y (\mathbf{r}) - y \tau_x (\mathbf{r}) \right] d^3 \mathbf{r}, \\
    t_x^{(\tau)} &= \frac{1}{\sqrt{2}} \left( w^{211}_{-1} - w^{211}_{1} \right) = \int \left[ y \tau_z (\mathbf{r}) - z \tau_y (\mathbf{r}) \right] d^3 \mathbf{r}. 
\end{align*}
Similarly, the totally symmetric, traceless, part reads 
\begin{equation}
    \widetilde{P}^{(T)} = \begin{pmatrix} \frac{1}{2} \left( q_{x^2-y^2}^{(\tau)} - q_{z^2}^{(\tau)} \right) & q_{xy}^{(\tau)} & q_{xz}^{(\tau)} \\ q_{xy}^{(\tau)} & - \frac{1}{2} \left( q_{x^2-y^2}^{(\tau)} + q_{z^2}^{(\tau)} \right) & q_{yz}^{(\tau)} \\ q_{xz}^{(\tau)} & q_{yz}^{(\tau)} & q_{z^2}^{(\tau)} \end{pmatrix},
\end{equation}
where we introduced the quadrupoles of the toroidization density, $q^{(\tau)}$, which correspond to $w^{2 1 2}$: 
\begin{align*}
    q_{xy}^{(\tau)} &= \frac{i}{\sqrt{2}} \left( w^{212}_{-2} - w^{212}_{2} \right) = \frac{1}{2} \int \left[ x \tau_y (\mathbf{r}) + y \tau_x (\mathbf{r}) \right] d^3 \mathbf{r}, \\
    q_{yz}^{(\tau)} &= \frac{i}{\sqrt{2}} \left( w^{212}_{-1} + w^{212}_{1} \right) = \frac{1}{2} \int \left[ y \tau_z (\mathbf{r}) + z \tau_y (\mathbf{r}) \right] d^3 \mathbf{r}, \\ 
    q_{z^2}^{(\tau)} &= w^{212}_{0} = \frac{1}{3} \int \left[ 2 z \tau_z (\mathbf{r}) - x \tau_x (\mathbf{r}) - y \tau_y (\mathbf{r}) \right] d^3 \mathbf{r}, \\
    q_{xz}^{(\tau)} &= \frac{1}{\sqrt{2}} \left( w^{212}_{-1} - w^{212}_{1} \right) = \frac{1}{2} \int \left[ x \tau_z (\mathbf{r}) + z \tau_x (\mathbf{r}) \right] d^3 \mathbf{r}, \\
    q_{x^2-y^2}^{(\tau)} &= \frac{1}{\sqrt{2}} \left( w^{212}_{-2} + w^{212}_{2} \right) = \frac{1}{2} \int \left[ x \tau_x (\mathbf{r}) - y \tau_y (\mathbf{r}) \right] d^3 \mathbf{r}.
\end{align*}
The trace part has to vanish since $\widetilde{P}$ is traceless: in the parallelism with the multipoles of the toroidal moment density, this can be explained because the trace would correspond to the monopole of the toroidal moment density, which is zero since $a^{(\tau)} = \mathbf{r} \cdot \boldsymbol{\tau} = \mathbf{r} \cdot (\mathbf{r} \times \boldsymbol{\mu}) = 0$.

$\widetilde{P}$ can be finally lifted to its rank-3 form $N^2$ using Eq. \eqref{eq14}. Below we write the two components $N^{2 \, (\mathbf{5})}$ and $N^{2 \, (\mathbf{3})}$ obtained from $\widetilde{P}^{(T)}$ and $\widetilde{P}^{(A)}$ and transforming as $\mathbf{5}$ and $\mathbf{3}$, respectively: 
\begin{widetext}
\begin{equation}
    \label{eq36}
    \begin{split}
        N^{2 \, (\mathbf{5})}_{1jk} &= \frac{1}{3} \begin{pmatrix} 0 & - q^{(\tau)}_{xz} & q^{(\tau)}_{xy} \\ - q^{(\tau)}_{xz} & -2 q^{(\tau)}_{yz} & - \frac{1}{2} \left( q^{(\tau)}_{x^2-y^2} + 3 q^{(\tau)}_{z^2} \right) \\ q^{(\tau)}_{xy} & - \frac{1}{2} \left( q^{(\tau)}_{x^2-y^2} + 3 q^{(\tau)}_{z^2} \right) & 2 q^{(\tau)}_{yz} \end{pmatrix}, \\
        N^{2 \, (\mathbf{5})}_{2jk} &= \frac{1}{3} \begin{pmatrix} 2 q^{(\tau)}_{xz} & q^{(\tau)}_{yz} & - \frac{1}{2} \left( q^{(\tau)}_{x^2-y^2} + 3 q^{(\tau)}_{z^2} \right) \\ q^{(\tau)}_{yz} & 0 & - q^{(\tau)}_{xy} \\ - \frac{1}{2} \left( q^{(\tau)}_{x^2-y^2} + 3 q^{(\tau)}_{z^2} \right) & - q^{(\tau)}_{xy} & -2 q^{(\tau)}_{xz} \end{pmatrix}, \\
        N^{2 \, (\mathbf{5})}_{3jk} &= \frac{1}{3} \begin{pmatrix} -2 q^{(\tau)}_{xy} & q^{(\tau)}_{x^2-y^2} & - q^{(\tau)}_{yz} \\ q^{(\tau)}_{x^2-y^2} & 2 q^{(\tau)}_{xy} & q^{(\tau)}_{xz} \\ - q^{(\tau)}_{yz} & q^{(\tau)}_{xz} & 0 \end{pmatrix},
    \end{split}
\end{equation}
\begin{equation}
    \label{eq37}
    \begin{split}
        N^{2 \, (\mathbf{3})}_{1jk} &= \frac{1}{3} \begin{pmatrix} 0 & t^{(\tau)}_{y} & t^{(\tau)}_{z} \\ t^{(\tau)}_{y} & -2 t^{(\tau)}_{x} & 0 \\ t^{(\tau)}_{z} & 0 & - 2 t^{(\tau)}_{x} \end{pmatrix}, \\
        N^{2 \, (\mathbf{3})}_{2jk} &= \frac{1}{3} \begin{pmatrix} - 2 t^{(\tau)}_{y} & t^{(\tau)}_{x} & 0 \\ t^{(\tau)}_{x} & 0 & t^{(\tau)}_{z} \\ 0 & t^{(\tau)}_{z} & -2 t^{(\tau)}_{y} \end{pmatrix}, \\
        N^{2 \, (\mathbf{3})}_{3jk} &= \frac{1}{3} \begin{pmatrix} -2 t^{(\tau)}_{z} & 0 & t^{(\tau)}_{x} \\ 0 & - 2 t^{(\tau)}_{z} & t^{(\tau)}_{y} \\ t^{(\tau)}_{x} & t^{(\tau)}_{y} & 0 \end{pmatrix}.
    \end{split}
\end{equation}
\end{widetext}

\section{A case study: \texorpdfstring{C\lowercase{r}$_2$O$_3$}{Cr2O3}}

\label{sec3}

Next, we illustrate the formalism developed in the previous Sections using first-principles density functional calculations for the prototypical linear magnetoelectric Cr$_2$O$_3$. While Cr$_2$O$_3$ is non-centrosymmetric and so has no net quadratic magnetoelectric response, we show that individual Cr$^{3+}$ ions possess local magnetic octupoles, in addition to their established local magnetoelectric multipoles. In contrast to the magnetoelectric multipoles, however, which have the same sign on each Cr ion, the magnetic octupoles are arranged in an {\it anti}-ferroic pattern. We then simulate the structural changes induced by an electric field by displacing individual Cr$^{3+}$ ions relative to their coordinating O$^{2-}$ ions and calculate the resulting change in local magnetic moment. We obtain a local second-order magnetoelectric response, with an overall anti-magnetoelectric pattern, consistent with the local magnetic octupoles. 

\subsection{Computational details}

\begin{figure}[t]
\includegraphics[width=0.42\textwidth]{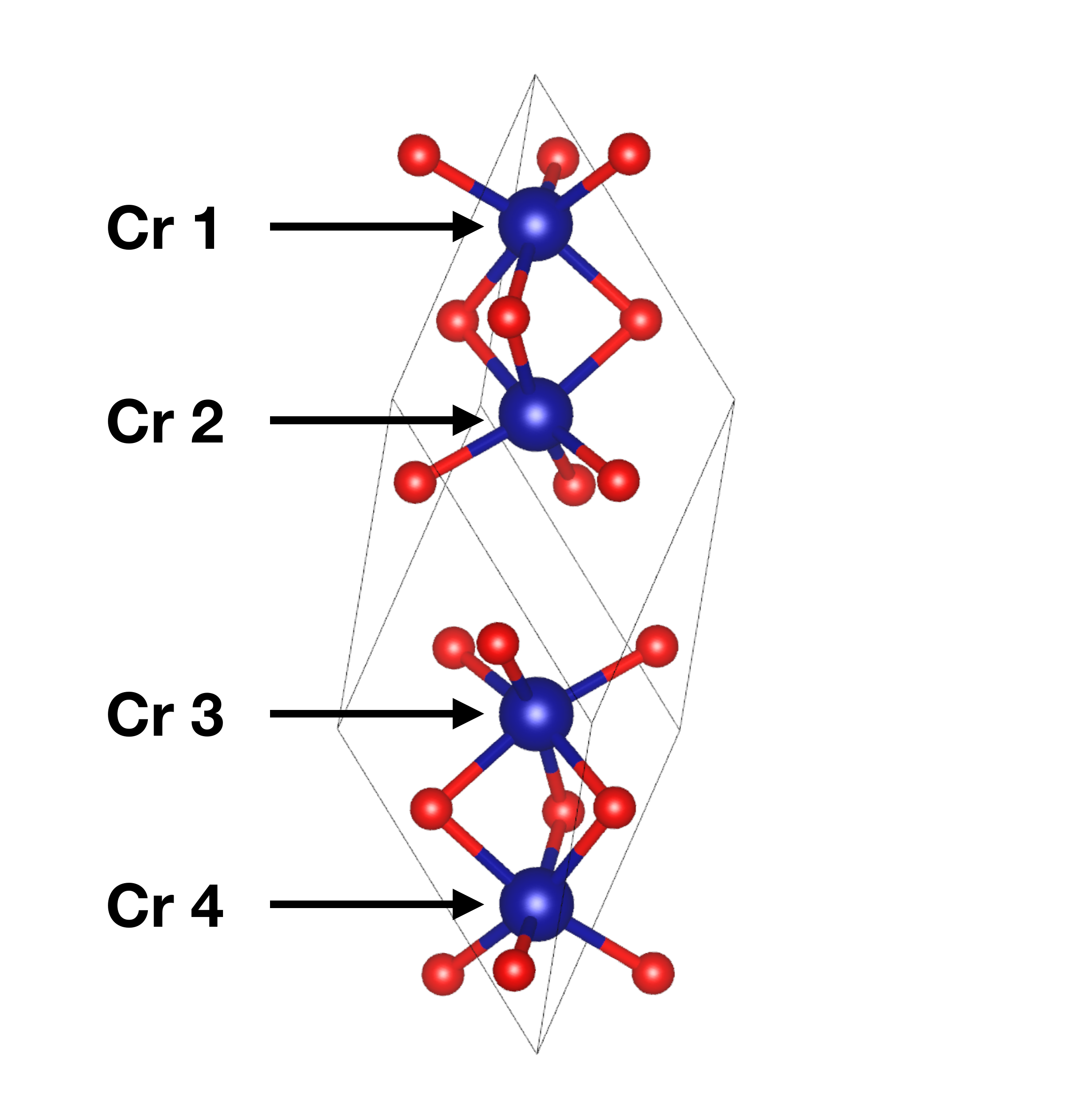}
\caption{\label{f1} Structure of Cr$_2$O$_3$. The Cr and O ions are shown in blue and red, respectively. The black lines outline the rhombohedral unit cell used in the calculations.}
\end{figure}

Our density functional calculations were performed within the local spin density approximation (LSDA) scheme \cite{LDA_pz}, as implemented in VASP \cite{VASP_1, VASP_2}. The atoms were described by scalar relativistic PAW pseudopotentials (PPs) \cite{pseudo_VASP, PAW_Blochl} with 3$p$, 4$s$, and 3$d$ valence electrons for Cr (PP \texttt{Cr$\_$sv}) and with 2$s$ and 2$p$ valence electrons for O (PP \texttt{O}). Correlation effects were dealt with by applying the rotationally invariant Hubbard U correction \cite{Dudarev_U} on Cr $d$ states, with U $= 4$ eV and J $= 0.5$ eV, which gives good agreement with experiments for the lattice parameters, the electronic band gap, and the magnetic moments on the Cr sites. Self-consistent calculations for the ground state and distorted structures, modeled with the primitive rhombohedral cell (magnetic space group R$\bar{3}'$c$'$), shown in Fig. \ref{f1}, were performed at the theoretical LSDA+U lattice constants $a = 5.312$ \AA, $\alpha = 54.87^{\circ}$, which are 0.85 \% and 0.23 \% smaller than the experimental values \cite{Cr2O3_exp_crystal} ($a_{\text{exp}} = 5.358$ \AA, $\alpha_{\text{exp}} = 55.0 ^{\circ}$), respectively. The pseudowave functions were expanded in a plane-wave basis set with a kinetic energy cut-off of 500 eV. The Brillouin Zone (BZ) was sampled using a uniform $\Gamma$-centered Monkhorst-Pack $\mathbf{k}$-point mesh \cite{Monkhorst_Pack_mesh} of $6 \times 6 \times 6$ points. The occupancies were treated with the tetrahedron smearing method with Bl\"ochl corrections. 

The irreducible spherical components of the magnetic octupole tensor, $w^{213}$, $w^{212}$, and $w^{211}$, were computed by decomposing the density matrix into spherical tensor moments \cite{multipole_decomposition}, as implemented in the FP-LAPW code \texttt{ELK} \cite{Elk_code}. Self-consistent calculations on the ground-state structure were performed at the aforementioned theoretical LSDA+U lattice constants. The APW functions were expanded in the spherical harmonic basis set, with a cut-off $l_{\text{max (apw)}} = 8$, and the BZ was sampled with a $6 \times 6 \times 6$ $\Gamma$-centered uniform Monkhorst-Pack mesh. 

\subsection{Discussion} 

As pointed out in Section \ref{intro}, the linear magnetoelectric response can be understood in terms of the multipole tensor $\mathcal{M}$ reported in Eq. \eqref{eq_me_tensor}. Cr$_2$O$_3$ has both a net monopole, $a$, and $z^2$ quadrupole, $q_{z^2}$, hence   
\begin{equation}
\label{eq39}
\mathcal{M} = \begin{pmatrix} a - \frac{1}{2} q_{z^2} & 0 & 0 \\ 0 & a - \frac{1}{2} q_{z^2} & 0 \\ 0 & 0 & a + q_{z^2}
\end{pmatrix}.
\end{equation}
In addition to this overall monopole and $z^2$ quadrupole, each Cr ion possesses its own local monopole and $z^2$ quadrupole, which are identical by symmetry on all the Cr sites (Table  \ref{t2}).
 The bulk linear magnetoelectric response then arises, in a simplified picture, from a superposition of the local magnetic responses induced by the ionic displacements due to an applied external electric field. Specifically, the local magnetic moment $\boldsymbol{\delta} \mathbf{m}$ linearly induced by a single ionic displacement $\mathbf{u}$ can be written as follows: 
\begin{equation}
    \delta m_i = \mathcal{Z}^m_{\text{loc}, ij} u_j \quad .
\end{equation}
Here, the local linear magnetic response given by the matrix $\mathcal{Z}^m_{\text{loc}}$ is a local analogue of $\mathcal{M}$, and contains contributions from all symmetry-allowed multipoles at the specific Wyckoff site. 

In Cr$_2$O$_3$, the symmetry of the Cr Wyckoff sites also allows for a local toroidal moment pointing along $z$, $t_z$. The local toroidal moments, however, have opposite signs on the two pairs of Cr ions in the unit cell (Table ~\ref{t2}),  and the local magnetoelectric multipole tensors at the Cr sites are 
\begin{equation}
    \begin{split}
\mathcal{M}_{\text{loc}} (\text{Cr1}, \text{Cr4}) &= \begin{pmatrix} a - \frac{1}{2} q_{z^2} & t_z & 0 \\ - t_z & a - \frac{1}{2} q_{z^2} & 0 \\ 0 & 0 & a + q_{z^2} \end{pmatrix}, \\
\mathcal{M}_{\text{loc}} (\text{Cr2}, \text{Cr3}) &= \begin{pmatrix} a - \frac{1}{2} q_{z^2} & - t_z & 0 \\ t_z & a - \frac{1}{2} q_{z^2} & 0 \\ 0 & 0 & a + q_{z^2} \end{pmatrix}, 
\end{split}
\end{equation}
where the signs of the entries in $\mathcal{M}_{\text{loc}}$ follow those of Table \ref{t2}. The ferroic order of $a$ and $q_{z^2}$ is consistent with the net bulk diagonal magnetoelectic response of Eq. \eqref{eq39}. In contrast the anti-ferroic arrangement of $t_z$ means that, while an in-plane off-diagonal $xy$ magnetoelectric response is allowed locally, the overall response is {\it anti}-magnetoelectric, and the bulk $xy$ (and $yx$) component of $\mathcal{M}$ and $\alpha$ vanishes. 

\begingroup
\begin{table}[t]
\centering
\begin{tabular}{|c|c|}
\hline
Magnetoelectric & Ordering \\
multipole & \\
\hline
\hline
$a$ & $+ + + +$ \\ 
\hline 
$q_{z^2}$ & $+ + + +$ \\
\hline
$t_z$ & $+ - - +$ \\ 
\hline 
\end{tabular}
\caption{Ordering of the symmetry-allowed magnetoelectric multipoles on the Cr atoms of Cr$_2$O$_3$.} 
\label{t2}
\end{table}
\endgroup

Extending the local magnetic response to second order in the local atomic displacements,  $\mathbf{u}$, we obtain
\begin{equation}
    \label{eq32}
    \delta m_i = \mathcal{Z}^m_{\text{loc}, i j} u_j + \mathcal{Z}^{m \, (2)}_{\text{loc}, i j k} u_j u_k.
\end{equation}
Here the rank-3 tensor $\mathcal{Z}^{m \, (2)}_{\text{loc}}$, describes the local second-order magnetoelectric response and has the form of the local magnetic octupole tensor. Likewise, the bulk second-order magnetoelectric response, introduced as $\beta_{ijk}$ in Section \ref{intro}, will be non-vanishing if the corresponding octupole components of the local $\mathcal{M}_{ijk}$ follow a ferroic order, but will be zero if they are anti-ferroically arranged. 
\begingroup
\begin{table}[t]
\centering
\begin{tabular}{|c|c|c|}
\hline
Multipole & Magnitude ($\mu_{\text{B}}$) & Ordering \\
\hline
\hline
$O_{-3}$ & 6.4 $\times 10^{-4}$ & $+ + - -$ \\ 
\hline 
$O_0$ & 1.1 $\times 10^{-2}$ & $- + - +$ \\ 
\hline 
$O_3$ & 2.4 $\times 10^{-4}$ & $- + - +$ \\
\hline 
$q_{z^2}^{(\tau)}$ & 1.9 $\times 10^{-3}$ & $+ + - -$ \\ 
\hline 
$t_{z}^{(\tau)}$ & 9.7 $\times 10^{-3}$ & $- + - +$ \\
\hline
\end{tabular}
\caption{Magnitude and pattern for the allowed components of the multipoles $w^{21r}$ ($r=1,2,3$) at the Cr sites.} 
\label{t4}
\end{table}
\endgroup
\begin{figure}[t]
\includegraphics[width=0.42\textwidth]{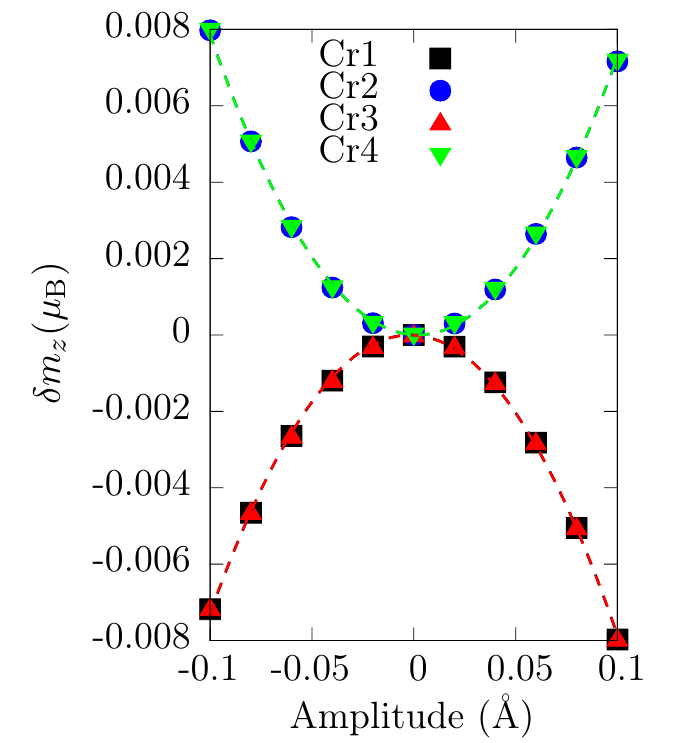}
\caption{\label{f2} z component of the change in the local Cr magnetic moment induced by a displacement of that Cr atom along x, as a function of the amplitude of the displacement. Cr atoms are labeled following Fig. \ref{f1}. Dashed lines represent the fitting curves.}
\end{figure}
\subsection{Results}

In Cr$_2$O$_3$ the irreducible spherical components of the octupole tensor allowed by symmetry are $O_{-3}$, $O_0$, $O_3$, $q^{(\tau)}_{z^2}$, and $t^{(\tau)}_z$. Their magnitudes and their relative arrangement in the unit cell, calculated using the \texttt{ELK} code, are reported in Table \ref{t4}. They all have an anti-ferroic arrangement, therefore a second-order \textit{anti}-magnetoelectric response is expected. 

To show more clearly the quadratic response, we choose the directions of $\mathbf{u}$ and $\boldsymbol{\delta} \mathbf{m}$ such that the linear contribution to $\boldsymbol{\delta} \mathbf{m}$ in Eq. \eqref{eq32} vanishes, thus making the second-order term  the leading one: in Cr$_2$O$_3$ this happens for the $z$ component of the local magnetic moment induced by a displacement of a Cr ion along $x$ or $y$, and for the $x$ and $y$ components of $\boldsymbol{\delta} \mathbf{m}$ induced by a displacement along $z$. As a proof of concept, we take the $z$ component of $\boldsymbol{\delta} \mathbf{m}$ induced by a displacement of each Cr ion, separately, along $x$. Following Eq. \eqref{eq32}: 
\begin{equation}
    \delta m_z = \mathcal{Z}^{m \, (2)}_{\text{loc}, zxx} u_x^2.
\end{equation}
In Fig. \ref{f2} we show our calculated change in local Cr magnetic moment, $\delta m_z$, as a function of the Cr ion displacement along the $x$ direction. We see that the response is purely quadratic as expected. In Table \ref{t3} we report the fitted values of $\mathcal{Z}^{m \, (2)}_{\text{loc}, zxx}$. They have the same magnitude but alternating signs, following the pattern of the corresponding components of $O$, $q^{(\tau)}$, and $\mathbf{t}^{(\tau)}$ in the $zxx$ entry of the irreducible spherical components of the octupole tensor, reported in Eqs. \eqref{eq35}, \eqref{eq36}, and \eqref{eq37}. The net quadratic magnetoelectric response is zero as required by symmetry, but our calculations reveal a hidden quadratic anti-magnetoelectric response that, to our knowledge, has not yet been observed experimentally. Among the other allowed responses, $\mathcal{Z}^{m \, (2)}_{\text{loc}, zyy}$ behaves in a similar way to $\mathcal{Z}^{m \, (2)}_{\text{loc}, zxx}$ discussed above, whereas the $\mathcal{Z}^{m \, (2)}_{\text{loc}, xzz}$ and $\mathcal{Z}^{m \, (2)}_{\text{loc}, yzz}$ responses are forbidden, because the corresponding irreducible components of the octupole tensor, $O_1$, $q^{(\tau)}_{yz}$, $q^{(\tau)}_{xz}$, $t^{(\tau)}_x$, and $t^{(\tau)}_y$, vanish.

\begingroup
\begin{table}[t]
\centering
\begin{tabular}{|c|c|}
\hline
Displaced atom & $\mathcal{Z}^{m \, (2)}_{zxx}$ ($\mu_{\text{B}} / \text{\AA}^2$)\\
\hline
\hline
Cr1 & -0.758 $\pm$ 0.008 \\ 
\hline 
Cr2 & 0.757 $\pm$ 0.008 \\ 
\hline 
Cr3 & -0.756 $\pm$ 0.008 \\
\hline 
Cr4 & 0.757 $\pm$ 0.008 \\
\hline
\end{tabular}
\caption{Fitted values for the second-order coefficient of the magnetic response along z induced by an ionic displacement along x.} 
\label{t3}
\end{table}
\endgroup

\section{Summary} 
In summary, we have demonstrated the relevance of the magnetic octupole tensor for describing the second-order magnetoelectric effect in materials. Using group theoretical analysis, we decomposed the octupole tensor into its irreducible spherical components: (i) a totally symmetric tensor, whose independent components are the magnetic octupoles transforming as the $l = 3$ irrep of SO(3), (ii) a trace tensor, and (iii) two residual tensors, whose independent parameters can be cast into multipoles of the toroidal moment density. 

We then showed that the irreducible components of the magnetic octupole tensor provide a useful tool for interpreting the second-order magnetoelectric response. In particular, we introduced the concept of {\it local} second-order magnetoelectric response, which describes the local magnetic response to ionic displacements, and showed that is captured by the local magnetic octupoles on the atomic sites. A ferroic ordering of these local octupoles results in a net second-order magnetoelectric response, whereas antiferroic ordering causes zero net response, but rather a hidden second-order anti-magnetoelectricity.

Finally, as an example, we considered the prototypical linear magnetoelectric Cr$_2$O$_3$. We used first-principles density functional theory to calculate the size and sign of the local octupoles on the Cr ions, as well as the magnetic response to ionic displacements. We found an antiferroic ordering of local Cr octupoles and corresponding anti-ferroic local second-order $zxx$ and $yxx$ magnetic responses. Therefore we predict a net second-order anti-magnetoelectric response in Cr$_2$O$_3$. In the spirit of Igor Dzyaloshinksii, we hope that our predictions stimulate experimental efforts to measure this intriguing hidden order in Cr$_2$O$_3$ or related materials. 

\label{sec4}

\section*{Acknowledgments}
The authors thank Dr. Alessio Caddeo for valuable discussions, and Dr. Sayantika Bhowal and Xanthe H. Verbeek for insightful comments. NAS and AU were supported by the ERC under the EU’s Horizon 2020 research and innovation programme grant No 810451 and by the ETH Z\"urich. Computational resources were provided by ETH Z\"urich’s Euler cluster.   

\appendix 
\section{Linear dependence of \texorpdfstring{$\widetilde{L}$, $\widetilde{P}$, $\widetilde{Q}$}{}}
\label{lin_dep}

We prove that $\widetilde{L}$, $\widetilde{P}$, and $\widetilde{Q}$ are linearly dependent. Starting from the definitions given in Eq. \eqref{eq31}, their sum reads 
\begin{equation}
\begin{split}
\label{eq16}
    \widetilde{L}_{ii'} + \widetilde{P}_{ii'} + \widetilde{Q}_{ii'} = &\left( M_{i j' k'} + M_{k' i j'} + M_{j' k' i} \right) \epsilon_{j'k'i'} \\ & - 6 \xi \delta_{ii'}.
\end{split}
\end{equation}
We further exploit the property $\epsilon_{j' k' i'} = - \epsilon_{k' j' i'}$, to rearrange the sum over $j'$ and $k'$ in this way: 
\begin{equation}
\label{eq14-2}
\begin{split}
    & \left( M_{i j' k'} + M_{k' i j'} + M_{j' k' i} \right) \epsilon_{j'k'i'} \\
    = & \, \frac{1}{2} \left( M_{i j' k'} - M_{i k' j'} + M_{k' i j'} - M_{j' i k'} + M_{j' k' i} \right. \\ & \left. - M_{k' j' i} \right) \epsilon_{j' k' i'}. 
    \end{split}
\end{equation}
The quantity in round brackets equals $6$ times the totally anti-symmetric component $A_{ij'k'} = \xi \epsilon_{ij'k'}$, where $\xi$ is defined as in Eq. \eqref{eq15}, hence 
\begin{equation}
\begin{split}
    \frac{1}{2} \left( M_{i j' k'} + M_{k' i j'} + M_{j' k' i} \right) \epsilon_{j'k'i'} &= 3 \xi \epsilon_{ij'k'} \epsilon_{j'k'i'} \\ 
    &= 6 \xi \delta_{ii'},
\end{split}
\end{equation}
where we used the property $\epsilon_{i j k} \epsilon_{i' j k} = 2 \delta_{i i'}$. As a consequence, the sum in Eq. \eqref{eq16} reads 
\begin{equation}
    \widetilde{L}_{ii'} + \widetilde{P}_{ii'} + \widetilde{Q}_{ii'} = 0.
\end{equation}
\begingroup
\begin{table*}[th]
\centering
\begin{tabular}{|c|c|c|}
\hline
Spherical harmonic ($l=3$) & Transformed starting & Permutations (multiplicity) \\
& expression & \\
\hline
\hline
$Y_{3 \, -3} \propto y (3 x^2 - y^2)$ & $\mu_y (3 x^2 - y^2)$ & $3 \mu_x x y -\mu_y y^2$ (2) \\
\hline
$Y_{3 \, -2} \propto x y z$ & $\mu_x y z$ & $\mu_y x z$ (1) \\ 
& & $\mu_z x y$ (1) \\
\hline
$Y_{3 \, -1} \propto y (4 z^2 - x^2 - y^2)$ & $\mu_y (4 z^2 - x^2 - y^2)$ & $4 \mu_z y z - \mu_x y x - \mu_y y^2$ (2) \\ 
\hline
$Y_{3 \, 0} \propto z (2 z^2 - 3 x^2 - 3 y^2)$ & $\mu_z (2 z^2 - 3 x^2 - 3 y^2)$ & $2 \mu_z z^2 - 3 \mu_x z x - 3 \mu_y z y$ (2) \\
\hline
$Y_{3 \, 1} \propto x (4 z^2 - x^2 - y^2)$ & $\mu_x (4 z^2 - x^2 - y^2)$ & $4 \mu_z x z - \mu_x x^2 - \mu_y x y$ (2) \\
\hline
$Y_{3 \, 2} \propto z (x^2 - y^2)$ & $\mu_z (x^2 - y^2)$ & $\mu_x z x - \mu_y z y$ (2) \\ 
\hline
$Y_{3 \, 3} \propto x (x^2 - 3 y^2)$ & $\mu_x (x^2 - 3 y^2)$ & $\mu_x x^2 - 3 \mu_y x y$ (2) \\
\hline
\end{tabular}
\caption{Intermediate steps used to build the magnetic octupoles. Left column: functional form of the spherical harmonics $Y_{l m_l}$ with $l = 3$. Center column: transformed expression to include one magnetic index. Right column: possible permutations and their multiplicity. A multiplicity equal to $2$ means that the two possible permutations give the same result.}
\label{t1}
\end{table*}
\endgroup

\section{Lifting the matrices \texorpdfstring{$\widetilde{L}$, $\widetilde{P}$, and $\widetilde{Q}$ to the residue tensor $N$}{}}
\label{residue}

To show how to lift the matrices $\widetilde{L}$, $\widetilde{P}$, and $\widetilde{Q}$ to rank-3 tensors, and derive Eq. \eqref{eq14}, we start from the definition of $\widetilde{L}$ and substitute the decomposition reported in Eq. \eqref{eq17}. We have: 
\begin{equation}
\label{eq20}
\begin{split}
    \widetilde{L}_{ii'} &= M_{i j' k'} \epsilon_{i' j' k'} - 2 \xi \delta_{i i'} \\
    &= S_{i j' k'} \epsilon_{i' j' k'} + \xi \epsilon_{i j' k'} \epsilon_{i' j' k'} + N_{i j' k'} \epsilon_{i' j' k'} - 2 \xi \delta_{ii'} \\
    &= N_{i j' k'} \epsilon_{i' j' k'},
\end{split}
\end{equation}
where we used the property of the Levi-Civita tensor $\epsilon_{i j k} \epsilon_{i' j k} = 2 \delta_{ii'}$, and the fact that the contraction of a totally symmetric tensor with $\epsilon$ is equal to $0$. Using the same arguments for the other independent matrix $\widetilde{P}$, we get: 
\begin{equation}
\label{eq21}
    \widetilde{P}_{ii'} = N_{k' i j'} \epsilon_{i' j' k'}.
\end{equation}
Eqs. \eqref{eq20} and \eqref{eq21} define a well-posed linear system, with $16$ equations (one for each independent component of $\widetilde{L}$ and $\widetilde{P}$) and $16$ unknowns (the $16$ independent components of $N$), therefore $N$ can be obtained by inverting such a linear system. This procedure would allow one to lift together (but not separately) $\widetilde{L}$ and $\widetilde{P}$ to a unique tensor $N$; with an additional step, the linear system can be further decoupled and $\widetilde{L}$ and $\widetilde{P}$ can be lifted separately. The first step to reduce $N$ is to write it in terms of two tensors $N^1$ and $N^2$, i.e.
\begin{equation}
    N_{ijk} = N^1_{ijk} + N^2_{ijk}.
\end{equation}
Such a reduction is not unique, and we have a freedom of choice on the definition of $N^1$ and $N^2$. The most convenient definition is such that the two tensors have the following symmetries by permutation of indices: 
\begin{align}
\label{eq29}
    N^1_{ijk} &= N^1_{kji}, \\
\label{eq30}
    N^2_{ijk} &= N^2_{ikj}, 
\end{align}
because in this way Eqs. \eqref{eq20} and \eqref{eq21} would read 
\begin{align}
\label{eq22}
    \widetilde{L}_{ii'} &= N^1_{i j' k'} \epsilon_{i' j' k'}, \\
\label{eq23}
    \widetilde{P}_{ii'} &= N^2_{k' i j'} \epsilon_{i' j' k'}.
\end{align}
Then the original linear system would be decoupled into two independent linear systems which separately link $\widetilde{L}$ and $\widetilde{P}$ to their rank-3 forms $N^1$ and $N^2$, respectively. 

We have an additional level of freedom in choosing how to decompose $N$, since we are free to choose which two of the three matrices $\widetilde{L}$, $\widetilde{P}$, and $\widetilde{Q}$ to consider as independent. As a consequence, the symmetry properties of $N^1$ and $N^2$ will be subject to the same freedom of choice. For instance, if we choose $\widetilde{P}$ and $\widetilde{Q}$ to be our independent matrices, $N^1$ and $N^2$ must have the following properties: 
    \begin{align}
        N^1_{ijk} &= N^1_{kji}, \\
        N^2_{ijk} &= N^2_{jik}
    \end{align}
    if we aim to write: 
    \begin{align}
        \widetilde{Q}_{ii'} &= N^1_{j' k' i} \epsilon_{i' j' k'}, \\
        \widetilde{P}_{ii'} &= N^2_{k' i j'} \epsilon_{i' j' k'}.
    \end{align}
    Likewise, if we choose $\widetilde{L}$ and $\widetilde{Q}$ as independent matrices, the symmetry properties of $N^1$ and $N^2$ are: 
    \begin{align}
        N^1_{ijk} &= N^1_{jik}, \\
        N^2_{ijk} &= N^2_{ikj}.
    \end{align}

To invert Eqs. \eqref{eq22} and \eqref{eq23} we note that an appropriate contraction with $\epsilon$ allows us to get a combination of entries of $N$. As an example, contraction of Eq. \eqref{eq22} with $\epsilon_{i'jk}$ gives: 
\begin{equation}
\begin{split}
    \widetilde{L}_{ii'} \epsilon_{i' j k} &= N^1_{i j' k'} \epsilon_{i' j' k'} \epsilon_{i' j k} \\
    &= N^1_{i j' k'} \left( \delta_{j' j} \delta_{k' k} - \delta_{j' k} \delta_{k' j} \right) \\ 
    &= N^1_{i j k} - N^1_{i k j}.
\end{split}
\end{equation}
To be able to isolate $N^1_{i j k}$ on the right-hand side, we write the most general possible combination of such contractions and require it to give $N^1_{i j k}$ as a result: 
\begin{equation}
    N^1_{i j k} = \alpha \widetilde{L}_{i i'} \epsilon_{i' j k} + \beta \widetilde{L}_{k i'} \epsilon_{i' i j} + \gamma \widetilde{L}_{j i'} \epsilon_{i' k i}.
\end{equation}
The property $N^1_{ijk} = N^1_{kji}$ implies $\gamma = 0$ and $\beta = - \alpha$. Furthermore, if we substitute $\widetilde{L}$ as given by Eq. \eqref{eq22}, we obtain 
\begin{equation}
\label{eq24}
\begin{split}
    N^1_{i j k} &= \alpha \left( \widetilde{L}_{i i'} \epsilon_{i' j k} - \widetilde{L}_{k i'} \epsilon_{i' i j} \right) \\
    &= \alpha \left( N^1_{i j' k'} \epsilon_{i' j' k'} \epsilon_{i' j k} - N^1_{k j' k'} \epsilon_{i' j' k'} \epsilon_{i' i j} \right) \\ 
    &= \alpha \left[ N^1_{i j' k'} \left( \delta_{j' j} \delta_{k' k} - \delta_{j' k} \delta_{k' j} \right) - N^1_{k j' k'} \left( \delta_{j' i} \delta_{k' j} \right. \right. \\ & \left. \left. \quad - \delta_{j' j} \delta_{k' i} \right) \right] \\ 
    &= \alpha \left( N^1_{i j k} - N^1_{i k j} - N^1_{k i j} + N^1_{k j i} \right) \\ 
    &= \alpha \left( 2 N^1_{i j k} - N^1_{j k i} - N^1_{k i j} \right) \\ 
    &= 3 \alpha N^1_{i j k}, 
\end{split}
\end{equation}
where in the third line we used the properties of $\epsilon$, in the second-to-last line we used the symmetry property of $N^1$, in particular $N^1_{k j i} = N^1_{i j k}$ and $N^1_{i k j} = N^1_{j k i}$, and in the last line we used the property of the residue tensor $N^1_{i j k} + N^1_{j k i} + N^1_{k i j} = 0$ (which can be proved by starting from the decomposition of $M$ reported in Eq. \eqref{eq17} and substituting the expressions for $S$ and $A$ given in Eqs. \eqref{eq13} and \eqref{eq38}, respectively). In order for Eq. \eqref{eq24} to hold, clearly $\alpha$ must be equal to $1/3$, hence we conclude that 
\begin{equation}
    N^1_{i j k} = \frac{1}{3} \left( \widetilde{L}_{i i'} \epsilon_{i' j k} - \widetilde{L}_{k i'} \epsilon_{i' i j} \right).
\end{equation}
Using the same arguments, $N^2$ can be written in terms of $\widetilde{P}$ as 
\begin{equation}
\label{eq33}
    N^2_{i j k} = - \frac{1}{3} \left( \widetilde{P}_{i i'} \epsilon_{i' j k} - \widetilde{P}_{j i'} \epsilon_{i' k i} \right).
\end{equation}
\section{Building the magnetic octupoles} 
\label{app_octupoles}

The magnetic octupoles introduced in Section \ref{octupoles}, Eq. \eqref{eq25}, can be built starting from the spherical harmonics with $l = 3$. However, in the present case one index of the magnetic octupole tensor $M$ describes the magnetization density rather than position, hence the spherical harmonics must be adapted accordingly. Starting from the functional form of a given spherical harmonic, one of the three spatial components must be transformed into a magnetic one: for example, $xyz$ becomes $\mu_x y z$, and $y x^2$ becomes $\mu_y x^2$. Finally, all the possible permutations must be considered and summed. As an example, we take the spherical harmonic $Y_{3 \, -2} = x y z$: we transform it into an object with one magnetic and two spatial indices, e.g. $\mu_x y z$, we consider all the possible permutations of indices, i.e. $\mu_x y z$, $\mu_y x z$, and $\mu_z x y$, and finally average them to get the expression of the octupole $O_{-2}$ reported in Eq. \eqref{eq25}: 
\begin{equation}
    O_{-2} = \frac{1}{3} \int \left[ \mu_x (\mathbf{r}) y z + \mu_y (\mathbf{r}) x z + \mu_z (\mathbf{r}) x y \right] d^3 \mathbf{r}.  
\end{equation}
By applying the same procedure for every $m_l = -3, -2 \dots, 2, 3 $, we get the expressions for all the remaining octupoles (Eq. \eqref{eq25}); we summarize the intermediate steps in Table \ref{t1}. 

\vspace{3cm}
\bibliographystyle{apsrev4-2}
\bibliography{references,references_2}

\end{document}